\documentstyle[12pt] {article}
\newcommand{\beq}{\begin{equation}}
\newcommand{\eeq}{\end{equation}}
\newcommand{\beqa}{\begin{eqnarray}}
\newcommand{\eeqa}{\end{eqnarray}}
\newcommand{\ba}{\begin{array}}
\newcommand{\ea}{\end{array}}

\begin{document}
\baselineskip 24 true pt

\def \mtil{{\tilde m}}

\begin{center}
{\bf ELECTRON CAPTURE RATES OF MID-fp SHELL NUCLEI FOR SUPERNOVA AND STELLAR EVOLUTION} \\
\end{center}
\vskip 0.5 truecm
\hoffset =-1.00 truecm
\begin{center}
{ S.Chakravarti$^{1}$,~K.Kar$^{2}$, A.Ray$^{3}$ and S.Sarkar$^{4}$ \\ \ \\

$^1$ Physics Department, Visva Bharati, Santiniketan 731235, India.\\
$^2$ Saha Institute of Nuclear Physics, 1/AF Bidhannagar, Calcutta
700064, India.\\

$^3$ Tata Institute of Fundamental Research, Homi Bhabha Road, Mumbai
400005, India. \\
$^4$ Ananda Mohan College, 102/1 Raja Rammohan Sarani, Calcutta 700009,
India. \\}

\end{center}
\textwidth=13.0 truecm
\vskip 0.5 truecm

{\bf Abstract:} A detailed model is constructed for the calculation of electron 
capture rates of
some $fp$ shell nuclei for situations prevailing in pre-supernova and
collapse phases of the evolution of the core of massive stars leading to
supernova explosion. The model uses explicitly the Gamow-Teller
strength function obtained through (n,p) reaction studies wherever available.
The rates include contribution from the excited states of the mother as well as
from the resonant states in equilibrium with the back reaction i.e. the beta
decay of the daughter nucleus. Comparisons are made with the shell model results
and the earlier calculations by Aufderheide $\it{et ~al}$. and 
Fuller, Fowler and Newman.
For the nuclei $^{56}$Fe, $^{55}$Mn and $^{60}$Ni with negative Q-values one observes
large contribution from the excited states.

\vfil\hfil\eject

  Weak interactions play a significant role in the evolution of massive stars
  at the presupernova stage as well as at the subsequent core collapse phase of
  type II supernovae. The competition between electron capture and beta decay
  on fp shell nuclei (mostly with ${50<A<70}$) at the late stages of silicon
  burning and the electron capture during the gravitational collapse, until
  neutrino trapping sets in, determines the final lepton fraction inside the
  core, which in turn determines the shock energy at core bounce. Thus a
  careful evaluation of the electron capture and beta decay rates at presupernova
  densities and the electron capture rates at higher densities during the
  core collapse are of utmost importance for the supernova problem. The
  earlier detailed model of Fuller, Fowler and Newman [FFN,1] had all the important
  physics aspects incorporated in it. Later it was conjectured [2,3] that
  some nuclei with ${A>60}$ not considered earlier may have significant impact
  on the deleptonisation and models for the calculations of the electron
  capture [4] and beta decay [5] rates for these nuclei were developed.
  Aufderheide et al [4] also identified sets of important nuclei for each
  density-temperature domain which need to be taken into account. Recently
  $0\hbar\omega$ shell model calculations in the full fp shell with all the
  4 orbits have been carried out for the evaluation of the electron capture
  and beta decay rates[6]. These rates are based on the allowed beta strength
  distributions using energy eigenvalues and wavefunctions obtained through
  very large matrix diagonalisation with a realistic fp shell interaction.

  On the other hand (p,n) reaction studies for a number of years and (n,p)
  reaction studies for the last decade or so carried out on a number of
  nuclei in this mass range have   given us valuable information on the same
  allowed beta decay and electron capture matrix elements. It is always
  observed that both for beta decay and electron capture the experimental
  strength distributions are quenched compared to the best theoretical
  estimates. The models take account of this by including a quenching
  factor in the Gamow-Teller (GT) strength distribution, mostly a constant
  independent of energy.

  In an earlier work Kar, Ray and Sarkar [5] had calculated the beta decay
  rates for a number of important nuclei with ${A>60}$ using a statistical
  model for the GT strength distribution including the effects of excited
  states and back resonances (coming from the equilibrium of beta decay and
  the inverse reaction of electron capture between the same two nuclei at
  typical high pre-supernova (preSN) temperatures). In this letter we extend that to
  the electron capture sector with the important improvement that the
  obseved (n,p) GT strength distribution is used directly for the evaluation
  of the rates for the ground state of the mother. So in this work we are
  concerned with the nuclei for which the (n,p) reaction results are
  available and we compare our results with those from  the shell model 
  calculations and other previous estimates.

  The electron capture rate on the nucleus $(Z,N)$ is given by

\begin{equation}
 {\lambda_0}^{EC}(\rho,T,Y_e) = ln~2 \sum_{i,j} {(2J_i+1) exp [-E_i/k_BT]
 \over G(Z,N,T)} {f_{ij}(\rho,T,Y_e,Q_{ij}) \over (ft)_{ij}}
\end{equation}

  Here $\rho$ is the matter density, $T$ the temperature, $k_B$ the Boltzmann
  constant and $Y_e$ is the electron fraction. $G(Z,N,T)$ is the nuclear
  partition function at temperature $T$, $E_i$ the excitation energy of the
  mother and $Q_{ij}$ and $(ft)_{ij}$ are the Q-value and the ft-value between
  the i-th state of the mother and the j-th state of the daughter.

\begin{equation}
      Q_{ij} = Q_{g.s.} + E_i - E_j
\end{equation}

  where $Q_{g.s.}$ is the ground state Q-value and $E_i$ is the excitation
  energy of the i-th excited state of the mother and $E_j$ is the corresponding
  energy of the j-th excited state of the daughter and 

\begin{equation}
	 (ft)_{ij} = {(6250s)\over (g_A/g_V)^2 B_{ij}(GT)} 
\end{equation}

as for the ground state and the low-lying states there is no contribution 
from the Fermi strength due to isospin selection rule. 
$B_{ij}(GT) = {\mid M_{ij}(GT_+)\mid}^2$ is the usual
nuclear matrix element squared of the Gamow-Teller ($GT_{+}$) operator.


  The phase space integral $f_{ij}$ is of the form [7]
  \begin{equation}
  f_{ij} = {1 \over {(m_{ec}^2)^5}} \int_{E_{min}}^{\infty} E (E^2 -
  m_e^2)^{1/2} (E-{\epsilon_0})^2 F_c(Z,E) FD(E-{\mu_e})
  FDC(E-{\mu_{\nu}}-{\epsilon_0}) dE
  \end{equation}
  with  $FD(x)= (1 + exp(x/k_BT))^{-1}$ and $FDC(x)=1-FD(x)$. $m_e$ refers to
  electron mass, ${\epsilon}_0=-Q_{ij}$ and the lower limit of the integral $E_{min}$
  is the larger value between ${\epsilon}_0$ and $m_e c^2$. For the Coulomb correction
  factor $F_c(Z,E)$ we use the form of ref[7].

      For PreSN temperatures the reverse reaction on the daughter
  nucleus, i.e. beta decay of the daughter happens to be in equilibrium with
  the electron capture. For the ground state or a lowlying
  state of the daughter the state of the mother that is in equilibrium is
  not the ground state or other lowlying states, but one that has maximum
  overlap with the daughter state differing from it by a particle-hole
  excitation lying a few MeV higher in energy.  We include the electron 
  capture rate from this specific resonant state separately and write
  the electron capture rate as,

  \begin{equation}
  \lambda^{EC}_{total}(\rho,T,Y_e) = \lambda^{EC}_{0}(\rho,T,Y_e) +
  \lambda^{EC}_{Res}(\rho,T,Y_e)
  \end{equation}

  where $\lambda^{EC}_{0}(\rho,T,Y_e)$ is given by eq (1).  $\lambda^{EC}
  _{Res}(\rho,T,Y_e)$, the resonance contribution calculated using the
  model of Fuller, Fowler and Newman [1] is given by, 


  \begin{eqnarray}
  \lambda^{EC}_{Res}&=&ln~2~(6250s)^{-1} (G^d /G^m)\times \hfil\nonumber \\
  &&\big[ ({g_A\over g_V})^2\times exp(-E_{Res}(GT)/K_BT) \times 
  {\mid M_{GT}\mid}^2 f(T,\mu,Q_{Res}(GT))  \nonumber \\ && +~ 
  exp(-E_{Res}(F)/K_BT)~\times {\mid M_{F}\mid}^2 f(T,\mu,Q_{Res}(F)\big]
  \end{eqnarray}

  where $G^d$ and $G^m$ are the partition functions of the daughter 
  and the
  mother, $E_{Res}(GT/F)$ is the 
  GT/F resonance energy in the mother
  and calculated using the model of FFN as described in ref [1].


 $Q_{Res}(GT/F) = Q_{g.s.} + E_{Res}(GT/F)$ for nuclei
 where the ground state single particle configurations of the daughter
 can get connected to the resonant state through GT/F transitions, 
 otherwise $Q_{Res}(GT/F) = Q_{g.s.}$. 
 ${\mid M_{GT} \mid}^2$ is the matrix element of Gamow-Teller operator 
 for the reverse reaction i.e. beta decay [1] and is given by,

  \begin{equation}
  {\mid M_{GT}\mid}^2 = Z_n \sum_{r,s} {\mid M^{s.p.}_{GT}(r,s)\mid}^2
  \times {{{n^r}_p {n^s}_h} \over(2j_s+1)}
  \end{equation}

  where  $n^r_n$  is the number of
neutron particles in orbit `r', $n^s_h$ is the number  of  proton
holes    in    orbit   `s'   (with   degeneracy   $2j_s+1$)   and
$M^{s.p.}_{GT}(r,s)$ is  the  single particle  GT  matrix  element
between  orbit  `r'  and  `s'. $Z_n$ is the quenching factor of the back
reaction (taken as 0.6). $ {\mid M_{F}\mid}^2 = (N-Z)+2$ i.e. the
difference in neutron and proton numbers of the daughter.   $E_{Res}(F)$ 
is the excitation  energy of  the  first  higher  isospin  state  
$(T_0+1)$  in  the mother ($T_0=(N-Z)/2$).

  For the Gamow-Teller strength distribution  needed in the 
  calculation of rates we use the results obtained from (n,p) experiments
  for the ground state of the mother. This is obtained by measuring the
  (n,p) reaction cross-section on the target and by performing multipole
  analyses on the resulting spectra. 
  These experimental strength distributions are known
  to be substantially quenched with respect to the theoretical predictions and 
  also the shape of the energy distribution is different compared to the theoretical forms.
  The theoretical forms of strength distribution depend on the model space 
  as well as the realistic interaction used.
  This is why the experimental strength distribution is the best form to use
  for the rates wherever available even though there are sometimes large (about
  20\%) uncertainty in the extraction of the strength.
  For the nuclei $^{56}$Fe [8], $^{55}$Mn [8] and $^{60}$Ni [9] considered here
  we use directly the histograms of the GT strength per MeV from (n,p) studies. 
  That is, we use the strength distribution as a function of energy instead of 
  ${\mid M_{ij}(GT_+)\mid}^2$ and replace the summation over the final states in
   eq(1) by integration over energy. 

  For the excited states the rates are calculated using Gaussian forms
  for the $GT_+$ strength distributions based on the spectral
  distribution theory arguments [10]. The same method was used in the
  beta decay sector in the model of Kar, Ray and Sarkar [5]. However as
  the excited states are within 3-4 MeV of excitation we use 
  the `Brink hypothesis'and take for the centroid and width of 
  the strength Gaussian the ground state values obtained from fits to
  (n,p) data. Thus the centroid of the strength Gaussian 
  corresponding to transition from the i-th excited state is 
  $E_C$(i) = $E_C$(g.s.) + $E_i$. The ground state strength 
  centroid is fixed by the Sutaria-Ray expression [11] which gives a global 
  fit to all known (n,p) data in fp shell as a function of $A$ and $N$. 
  For the widths we use best fit values obtained by Sutaria-Ray for the 
  specific examples [12].

  Table 1 gives the calculated total rates for the nuclei $^{56}$Fe, 
  $^{55}$Mn and $^{60}$Ni with the four representative density grid 
  points $\log~\rho_{10} = $-2.5, -1.5, -0.5 and 0.5 ($\rho_{10}$ is the 
  density in $10^{10} g/cc$)
  and five temperatures $T_9$ 
  (in $10^9$ K) =  2,3,4,5 and 6. The rates within parenthesis are 
  the ground state contributions. The quenching factor in our theoretical 
  model is taken as a parameter and fixed by the following procedure: 
  For high density 
   (log~$\rho_{10} = $0.5)
   and low temperatures ($T_9=$2) when the electron chemical potential is large 
   to see the full GT strength distribution the quenching factor is adjusted 
   to give the ground state rate identical to the one obtained by using 
   the (n,p) data. This quenching factor is then used for all the 
   excited states considered. The rates for all the three nuclei considered 
   ( all with negative Q-value for e-capture) show a very rapid 
  fall with decreasing density. This is because though the chemical potential
  ($\mu_e$) of the electrons is 11.98 MeV at log~$\rho_{10}=$0.5 
  ($T_9$=2,$Y_e$ =0.45) it becomes very small i.e. 0.804 MeV at 
  log~$\rho_{10}=$-2.5 ($T_9$=2,$Y_e$ =0.45) and at this lower density only 
  a very small fraction of the electrons from the tail of the Fermi-Dirac 
  distribution can overcome the Q-value and cause the capture. We also 
  observe that for the highest density tabulated the contribution
  from the excited states is small (at most about 30\% for the 
  highest temperature). But at the lower densities when $\mu_e$ is too small 
  to cause capture on the ground state, the excited states still 
  have a better chance for capture with Q-values higher (and sometimes even 
  positive) and often give dominant contribution. 
  On the other hand for the back resonances considered though both 
  Fermi and Gamow-Teller can in principle contribute , we find that 
  their contributions even at log~$\rho_{10} = $0.5 is essentially 
  zero (0.10 and 0.00
  for GT and Fermi respectively for $ ^{60}$Ni at $T_9=$6). 
  For the nuclei $^{56}$Fe, $^{55}$Mn and $^{60}$Ni the GT back resonances 
  are at excitations of 5.38, 6.71 and 5.29 MeV respectively (according to 
  the FFN prescription that we use) and these are too high to make any 
  significant contribution to the rates. The back resonance
  excitation energies for the Fermi are all higher than 10 MeV and here again 
  the exponential factor in eq (6) makes the contributions zero.

  In Table 2 we compare our rates with the rates calculated by other models.
  We find that for the two even-even nuclei our rates are more than an order of
  magnitude higher than the ones by Aufdeheide $\it{et~ al.}$ [4]. Actually 
  Aufderheide $\it{et~ al.}$ puts the total strength 
  in a single excited state of the daughter nucleus
  with an effective `log ft' and ref [6] observes that for the even-even nuclei
  it is put at too high an excitation. For the odd-A nucleus $^{55}$Mn however 
  our values are within a factor of 2 higher compared to Aufderheide 
  $\it{et~ al.}$. Compared to FFN our values are within a factor of 2 on the higher side except
  for low density ($\rho=5.86 \times 10^7$g/cc for $^{60}$Ni) where the 
  FFN value is an order of magnitude  lower. But FFN used phenomenological 
  log ft's concentrated in a giant resonance at a fixed energy whereas 
  we use the observed (n,p) strength distribution. For $^{56}$Fe 
  at a density of $10^8$g/cc our ground state 
 rate compares well with that of Martinez-Pinedo $\it{et~ al.}$[6] but 
 our total rate is
 higher by a factor of seven. A more detailed comparison for other densities and 
 other nuclei is needed to understand this.

In conclusion we stress that for a proper comparison of the electron
capture rates by different methods a calculation which uses observed
experimental ground state GT strength distributions is important and is 
reported in this letter.
 We are in the process of extending our calculations to all the nuclei for which
 (n,p) results are available. We also plan to look at the problem of
 identifying the sets of important nuclei for the preSN evolution for 
 different densities and 
 temperatures using our rates and the nuclear statistical equilibrium. 
 This may show important differences from 
 the ones prescribed by Aufderheide $\it{et~ al.}$ [4].

\vfil\eject

  {\bf References}

  \noindent [1] G.M.Fuller, W.A.Fowler and M.J.Newman, Ap.J. Sppl. 42 (1982) 447;
  48 (1982) 279; Ap.J. 252 (1982) 715; 293 (1985) 1 

  \noindent [2] G.E.Brown, Proc. Intl. Nucl. Phys. Conf.1989, eds M.S.Hussein et al
  (World Scientific, Singapore, 1990) p3
  
  \noindent [3] H.A.Bethe, Rev. Mod. Phys. 62(1990)801

  \noindent [4] M.B.Aufderheide, G.E.Brown, T.T.S.Kuo, D.B.Stout and P.Vogel, Ap. J.
  362 (199) 241; M.B. Aufderheide, I. Fushiki, S.E. Woosley and D.H. Hartmann,
  Ap.J. Suppl. 91 (1994) 389

  \noindent [5] K.Kar, A.Ray and S.Sarkar, Ap.J. 434(1994)662 

  \noindent [6] G.Martinez-Pinedo, K.Langanke and D.J. Dean, preprint nucl-th/9811095; 
     E.Caurier, K.Langanke, G.Martinez-Pinedo, and F.Nowacki, Nucl.  Phys. 653(1999)439. 

  \noindent [7] M.J. Murphy, Ap.J. Suppl. 42(1982)420

  \noindent [8] S. El-Kateb ${\it{et~ al}}$, Phys. Rev. C49(1994)3128

  \noindent [9] A.L. Williams ${\it{et~ al}}$, Phys. Rev. C51(1995)1144

  \noindent [10]   J.B. French and V.K.B. Kota Ann. Rev. Nucl. Particle
   Sc. 12 (1982) 35; V.K.B. Kota and K. Kar, 1988 Lab Rep UR - 1058 Univ
   Rochester; Pramana 32 (1989) 647 
   
  \noindent [11] F.K.Sutaria and A.Ray, Phys. Rev. C52(1995)3460  

  \noindent [12] F.K. Sutaria and A.Ray, 1997(unpublished) 

\vfil\eject


\begin{center}\bf TABLE 1 \end{center}
\medskip
\begin{center}{\bf e$^{-}$ Capture Rates for $Y_e$ = 0.45}\end{center}
\begin{center}
\begin{tabular}{|c|c|c|c|c|c|}\hline  
Nucleus & Temperature &\multicolumn{4}{c|}{$\log \rho_{10}$} \\   
\cline{3-6}
 & $^0K$ & 0.5 & -0.5 & -1.5 & -2.5\\ 
\hline
 &       &     &      &      & \\ 
 & $2\times 10^9$ & $231.1$ & $1.31\times 10^{-1}$ &
   $1.65\times 10^{-7}$ & $8.21\times 10^{-11}$\\ 
 &                & $(227.7)$ & $(1.22\times 10^{-1})$ &
 $(3.72\times 10^{-8})$  & $(1.13\times 10^{-11})$  \\
 & & & & & \\
 & $3\times 10^9$ &  $245.8$ & $2.06\times 10^{-1}$& $2.03\times 10^{-5}$& $1.14\times 10^{-7}$ \\
&  & $(229.6)$ & $(1.55\times 10^{-1})$& $(3.36\times 10^{-6})$& $(1.06\times 10^{-8})$ \\
& & & & & \\
  $^{56}$Fe &  $4\times 10^9$ &  $268.0$ & $3.55\times 10^{-1}$ & $3.54\times 10^{-4}$ & $5.76\times 10^{-6}$ \\
 & &   $(232.6)$ & $(2.08\times 10^{-1})$ & $(4.52\times 10^{-5})$ &
$(4.92\times 10^{-7})$ \\
& & & & & \\
 & $5\times 10^9$ & $298.5$ & $6.58\times 10^{-1}$ & $2.41\times 10^{-3}$ & $7.00\times 10^{-5}$\\
 & & $(236.6)$ & $(2.84\times 10^{-1})$ & $(2.63\times 10^{-4})$ & 
$(5.87\times 10^{-6})$ \\
& & & & & \\
 & $6\times 10^9$& $340.5$& $1.22\times 10^{0}$& $9.45\times 10^{-3}$& 
$4.13\times 10^{-4}$ \\
 & & $(241.7)$& $(3.19\times 10^{-1})$& $(9.73\times 10^{-4})$& $(3.57\times 10^{-5})$\\
& & & & & \\
\hline
& & & & & \\
 & $2\times 10^9$ & $110.6$ & $9.85\times 10^{-2}$ & $3.13\times 10^{-6}$ & 
$1.46\times 10^{-9} $  \\
 & & $(108.1)$ & $(8.03\times 10^{-2})$ & $(1.61\times 10^{-6})$ & 
$(4.00\times 10^{-10})$  \\
 & & & & & \\
 & $3\times 10^9$ & $114.3$ & $1.28\times 10^{-1}$& $4.84\times 10^{-5}$& $2.42\times 10^{-7}$\\
 & & $(109.4)$& $(9.11\times 10^{-2})$& $(1.68\times 10^{-5})$&
$(5.41\times 10^{-8})$ \\
 & & & & & \\
 $^{55}$Mn& $4\times 10^9$ & $121.4$ & $1.92\times 10^{-1}$ & $3.04\times 10^{-4}$ & $4.40\times 10^{-6}$ \\
&
&
$(111.3)$ &
$(1.09\times 10^{-1})$ &
$(7.76\times 10^{-5})$ &
$(8.54\times 10^{-7})$
\\
 & & & & & \\
 
& 
$5\times 10^9$ & 
$131.8$ & 
$3.02\times 10^{-1}$ & 
$1.15\times 10^{-3}$ & 
$3.05\times 10^{-5}$
\\
 
& 
& 
$(113.8)$ & 
$(1.36\times 10^{-1})$ &
$(2.48\times 10^{-4})$ & 
$(5.56\times 10^{-6})$
\\
 & & & & & \\

& 
$6\times 10^9$&
$144.4$&
$4.62\times 10^{-1}$&
$3.17\times 10^{-3}$&
$1.31\times 10^{-4}$
\\

&
&
$(116.9)$&
$(1.76\times 10^{-1})$&
$(6.57\times 10^{-4})$&
$(2.51\times 10^{-5})$
\\
 & & & & & \\
\hline\end{tabular}
\newpage
\begin{center}\bf TABLE 1 (contd.)\end{center}
\medskip
\begin{center}{\bf e$^{-}$ Capture Rates for $Y_e$ = 0.45}\end{center}
\begin{tabular}{|c|c|c|c|c|c|}
\hline  
Nucleus & Temperature & \multicolumn{4}{c|}{$\log\rho_{10}$} \\ 
\cline{3-6} 
        & $^0K$       & 0.5 & -0.5 & -1.5 & -2.5\\ \hline 
 & & & & & \\ 

&
$2\times 10^9$ &
$416.1$ & 
$6.66\times 10^{-1}$ &
$7.84\times 10^{-6}$ & 
$4.99\times 10^{-9}$ 
\\

& 
&
$(415.5)$ & 
$(6.61\times 10^{-1})$ & 
$(3.66\times 10^{-6})$ & 
$(9.64\times 10^{-10})$ 
\\
 & & & & & \\
 & 
$3\times 10^9$ &
$426.7$ &
$8.27\times 10^{-1}$&
$2.51\times 10^{-4}$&
$1.66\times 10^{-6}$
\\

&
&
$(418.7)$&
$(7.56\times 10^{-1})$&
$(6.84\times 10^{-5})$&
$(2.19\times 10^{-7})$
\\
 & & & & & \\

$^{60}$Ni&   
$4\times 10^9$ &
$457.3$ &
$1.25\times 10^{0}$ &
$2.45\times 10^{-3}$ &
$4.40\times 10^{-5}$
\\

&
&
$(423.3)$ &
$(8.96\times 10^{-1})$ &
$(4.34\times 10^{-4})$ &
$(4.74\times 10^{-6})$
\\
 & & & & & \\
 
& 
$5\times 10^9$ & 
$514.8$ & 
$2.14\times 10^{0}$ & 
$1.18\times 10^{-2}$ & 
$3.65\times 10^{-4}$
\\

& 
& 
$(429.4)$ & 
$(1.09\times 10^{0})$ &
$(1.64\times 10^{-3})$ & 
$(3.67\times 10^{-5})$
\\
 & & & & & \\

& 
$6\times 10^9$&
$593.2$&
$3.45\times 10^{0}$&
$3.55\times 10^{-2}$&
$1.62\times 10^{-2}$
\\

&
&
$(437.1)$&
$(1.35\times 10^{0})$&
$(4.58\times 10^{-3})$&
$(1.69\times 10^{-4})$
\\
 & & & & & \\
\hline
\end{tabular}
\end{center}
\vfil\eject



\hoffset -0.75in

\begin{center}\bf TABLE 2 \end{center}
\medskip
\begin{center}\bf Comparison of our electron capture rates with other model calculations\end{center}
\medskip
\begin{center}
\begin{tabular}{|c|c|c|c|c|c|c|c|}
\hline
Nucleus&Density&Temperature&$Y_e$&Our&\multicolumn{3}{c|}{Rates ($s^{-1}$)
 of}\\
 & ($\times$ 10$^7$ g/cc)&($\times$ 10$^9$ $^o$K)&&Rate&\multicolumn{3}{c|}{  }\\
\cline{6-8} 
&&&&($s^{-1}$)&Aufderheide & FFN&Martinez-Pinedo\\
&&&&          &{\it et al.}[4]&[1]&{\it et al.}[6]\\
\hline
&5.86&3.40&0.47&2.07E-06&6.97E-08&-&-\\
$^{56}$Fe&10.7&3.65&0.455&1.42E-05&4.68E-07&1.0E-05&2.1E-06\\
         &14.5&3.80&0.45 &3.97E-05&1.31E-06&2.81E-05&-     \\
\hline
&5.86&3.40&0.47&2.17E-05&1.49E-06&1.09E-03&-\\
$^{60}$Ni&10.7&3.65&0.455&1.25E-04&7.64E-06&-&-\\
         &14.5&3.80&0.45 &3.18E-04&2.74E-05&1.39E-04&-\\
         &33.0&4.24&0.44 &3.93E-03&3.34E-04&- &-\\
\hline
         &10.7&3.65&0.455&1.52E-05&9.23E-06&-&-\\
$^{55}$Mn&14.5&3.80&0.45 &3.79E-05&2.25E-05&2.03E-05&-\\
         &33.0&4.24&0.44 &4.55E-04&2.64E-04&- &- \\    
\hline
\end{tabular}
\end{center}

\end{document}